# Clustering Bioactive Molecules in 3D Chemical Space with Unsupervised Deep Learning

Chu Qin[†,⊥], Ying Tan [‡,⊥], Shang Ying Chen[†], Xian Zeng[†], Xingxing Qi[‡], Tian Jin[‡], Huan Shi[‡], Yiwei Wang[¶], Yu Chen[‡], Jingfeng Li[‡], Weidong He[†], Yali Wang[†], Peng Zhang[†], Feng Zhu[§], Hongping Zhao[#], Yuyang Jiang[‡,*], and Yuzong Chen[†,‡,*]

[†]BIDD Group Department of Pharmacy, National University of Singapore, 18 Science Drive 4, Singapore 117543, Singapore

[‡] The State Key Laboratory of Chemical Oncogenomics, Key Laboratory of Chemical Biology, the Graduate School at Shenzhen, Tsinghua University; Shenzhen Technology and Engineering Laboratory for Personalized Cancer Diagnostics and Therapeutics, Shenzhen Kivita Innovative Drug Discovery Institute, Shenzhen 518055, P. R. China

[§]Drug Research and Bioinformatics Group, College of Pharmaceutical Sciences, Zhejiang University, Hangzhou 310058, P. R. China

[#]School of Science, China Pharmaceutical University, Nanjing 210009, P. R. China

[¶]Shanghai Kudai Network Technology Co., Ltd. Shanghai, Shanghai 200000, P. R. China

**Corresponding Author**

*jiangyy@sz.tsinghua.edu.cn.

*phacyz@nus.edu.sg.

**Author Contributions**

[⊥]C.Q. and Y.T. contributed equally.

# ABSTRACT

Unsupervised clustering has broad applications in data stratification, pattern investigation and new discovery beyond existing knowledge. In particular, clustering of bioactive molecules facilitates chemical space mapping, structure-activity studies, and drug discovery. These tasks, conventionally conducted by similarity-based methods, are complicated by data complexity and diversity. We explored the superior learning capability of deep autoencoders for unsupervised clustering of 1.39 million bioactive molecules into band-clusters in a 3-dimensional latent chemical space. These band-clusters, displayed by a space-navigation simulation software, band molecules of selected bioactivity classes into individual band-clusters possessing unique sets of common sub-structural features beyond structural similarity. These sub-structural features form the frameworks of the literature-reported pharmacophores and privileged fragments. Within each band-cluster, molecules are further banded into selected sub-regions with respect to their bioactivity target, sub-structural features and molecular scaffolds. Our method is potentially applicable for big data clustering tasks of different fields.

**AUTHOR CONTRIBUTIONS**

Y.Z.C. conceived and directed the work, Y.Y.J. added ideas, C.Q. performed main computations and analysis, Y.T. and Y.W.W. developed and tested visualization software, S.Y.C. analyzed the results, F.Z. and P.Z. processed molecular data, X.Z., W.D.H. and Y.L.W. collected and analyzed pharmacophore and privileged structure data, X.X.Q., T.J. and X.Y.S. drew and analyzed colored molecular structure figures. H.P.Z. designed and tested computational procedures, Y.Z.C. drafted the paper. All authors read, edited and approved the final paper.

Unsupervised data clustering has found extensive applications in various data stratification, pattern categorization, and knowledge discovery tasks beyond existing knowledge(1). An outstanding problem is the clustering of big data beyond conventional similarity-based approaches (2-5). Specifically, unsupervised clustering of bioactive molecules with respect to their common molecular determinants(6-10) facilitates the mapping of bioactive or pharmacological chemical space, the investigation of structure-activity relationships, and drug discovery(10, 11). Several similarity-based methods have been employed for unsupervised grouping of molecules by hierarchical clustering of the molecular scaffolds(6) and substructure fingerprints(9), and by matching molecular fragments(7, 12), substructures(8, 12) and physicochemical properties(10). These methods employ the similarity principle for clustering molecules under the conventional molecular representations (molecular scaffolds, fragments, substructure fingerprints, and physicochemical properties). However, not all complex features of bioactivities are capturable by similarity-based analysis within conventional molecular frameworks. A new approach is needed for clustering molecules and other big data beyond the conventional algorithms and data representations.

A potentially-useful approach for unsupervised clustering beyond the conventional algorithms and representations is deep neural networks (DNNs). DNNs have been employed for unsupervised learning of the informative features by its unique ability to flatter the complex data(13) and segment them into classes of specific characteristics(14) while preserving the local and global data characteristics(15). The DNN-learned informative features have been used as pre-trained representation features of the conventional machine learning methods for improved data clustering(16) and classification(17). DNNs trained on 4.9-50 million samples have successfully accomplished difficult tasks(18, 19). These distinguished capabilities of DNNs in learning complex features from big data may be explored for unsupervised clustering of big data beyond the conventional frameworks.

DNN-learned informative features has been presented or projected to 2-dimensional latent space for revealing the data landscapes in that space(20-22). In some cases, the DNN-learned features in the 2-dimensional latent space present a landscape of band-clusters (data distributed into bands of particle streams), with each band-cluster grouping together subsets of the data of common frameworks but varying global features (e.g. images of a handwritten digit, Reuter newswire stories of a topic such as energy markets, and drug-like molecules of similar logP values)(20, 21, 23).

Hence, DNNs may have the ability to group data into band-clusters of common frameworks beyond similarity and conventional data representations. A question is whether this ability may be exploited for unsupervised clustering of structurally-diverse bioactive molecules into band-clusters of molecules of selected bioactivities with unique elements of pharmacophores(24), privileged substructures(25-27), and scaffold branches(11). Although the band-cluster characteristics are contained in the DNN-learned features, feeding these features into a conventional clustering method(16) subject them to the limitations of the similarity framework. Projection of the DNN-learned features to 2-dimentional space(20-22) lead to information loss in low-dimensionality and in some cases subject to the limitations of the dimensionality-reduction methods (e.g. PCA and t-SNE).

One alternative approach to alleviate these problems is to employ DNNs to directly learn informative features in a 3-dimentional latent space, wherein the DNN-generated band-clusters are subject to less information-loss than in 2-dimensional latent space, and the landscapes can be straightforwardly displayed and analyzed by a space-navigation simulation software without using a dimensionality-reduction method. A critical question is whether the DNN-learned features in 3-dimentional latent space can adequately indicate the band-clusters of common frameworks. This question was interrogated by analyzing the members of individual band-cluster with respect to their bioactivity-classes (e.g. kinase inhibitors) and the common sub-structural features with respect to the literature-reported pharmacophores(28) and privileged substructures(25, 26) of the bioactivity classes in the band-cluster.

We developed the deep autoencoders (DAEs) for unsupervised clustering of 1.39 million ChEMBL bioactive molecules(29) in a 3-dimensional latent chemical space, together with a chemical space navigation simulation software DeepChemScape for displaying and analyzing the band-cluster landscapes. The molecules of individual band-cluster were extracted by density analysis or by reference to the manually-selected molecules that visibly define band-cluster boundaries. The intra-cluster molecules of two selected band-clusters were analyzed for their bioactivity profiles and common sub-structural features in comparison with the literature-reported pharmacophores(28) and privileged structures(25, 26) of the bioactivities enriched in these band-clusters.

**Data collection, processing and molecular representation**

The SMILES structures, macromolecule targets and the activity values (in IC50, EC50, and Ki) of 1.7 million bioactive molecules were from the ChEMBL database version-22(29). The SMILES structures were processed by using Open Babel(30) to convert them into the MOL 2D format, remove single-atom structures and salts, discard the smaller independent structures, and add explicit hydrogens. There are 386,655 molecules active against 2,245 targets with IC50, EC50, or Ki <10 μM, a widely-used bioactivity cut-off(31). These 2,245 targets were mapped to 30 level-2 ChEMBL target families(29) (Supplementary Table S1), 1,398 Pfam domain families, or 2,251 Interpro domain families via CheMBL cross-links to the corresponding Uniprot and Interpro entries. There are 1,011,600 additional molecules with lower activities (IC50, EC50, or Ki >10 μM), ambiguous activity values (e.g. % inhibition at one concentration), or un-defined targets (e.g. cell-lines). These molecules of low or unspecified activities were used for training the DAE models.

In chemical space mapping(11), visualization(32) and virtual screening(33), molecules have been frequently represented by molecular fingerprints(9, 33) that encode molecular structures by a series of binary digits indicating the presence or absence of individual substructures(9, 33). Specifically, we used Padel(34) to compute the 881-bit Pubchem fingerprints for 1,623,663 molecules. These molecules were further divided into two sets: Version-19 set (1,398,255 molecules) and the new-additions set (225,408 molecules added after Version 19) for training and validating the DAE models.

**Unsupervised clustering and display of molecules in a 3-dimensional latent chemical space with deep autoencoder networks**

Our DAEs are composed of a pair of complementary DNNs. An auto-encoder converts molecular fingerprints into 3-imensional latent codes for data clustering. An auto-decoder converts the latent codes back to the original molecular fingerprints for optimizing the auto-encoder. The auto-encoder was trained and tested on 4/5 and 1/5 randomly selected molecules of the Version-19 set, and validated on the new-additions set. The optimization of the DAE hyperparameters was conducted in two phases as described in Supplementary Methods, which proceeds until the reconstruction rate reaches optimal value and the molecules are distributed in band-clusters(20, 21, 35) in the 3-dimensional latent chemical space as revealed by a chemical space navigation software DeepChemScape (Supplementary Methods).

**The band-clusters of 1.39 million bioactive molecules in 3-dimensional latent chemical space and the banded distributions of bioactivity classes**

Figure 1 shows the DAE-learned band-cluster landscape of the 1.39 million Chembl bioactive molecules in the 3-dimentional latent chemical space displayed on DeepChemScape. Molecules are represented as individual spheres in default grey colors except those in 10 selected band-clusters highlighted by other colors. Consistent with the literature-reported DNN-learned landscapes of various data-types in two-dimensional latent space(20, 21, 23), all molecules are aligned into band-clusters confined within a vast cone-shaped subspace. Band-clusters roughly originate from a common central region (center of the bioactive universe) and extend outwardly like straight particle streams. Many band-clusters are populated by the moderate-to-high (<10μM) potency molecules of selected bioactivity classes. Inspection of the ChEMBL level-2 target families(29) of these molecules revealed that each band-cluster typically groups together molecules of selected bioactivity classes. Figure 2 shows three band-clusters A, B and C with different bioactivity classes highlighted by different colors. The structure, target and molecular scaffolds of all molecules in band-cluster A and B are provided in Supplementary Table S2 and S3 respectively.

Band-cluster A (6121 molecules) is primarily composed of 67.55% molecules of low or unspecified activities, 10.50% family-A GPCR ligands, 8.66% kinase inhibitors, 2.6% nuclear receptor ligands, 2.5% protease inhibitors, 1.88% reader, 1.39% family-B GPCR ligands and 1.14% eraser inhibitors. Band-cluster B (3410 molecules) consists of 67.62% molecules of low or unspecified activities, 15.95% family-A GPCR ligands, 6.48% protease inhibitors, 2.35% voltage-gated ion channel blockers, 1.79% reader inhibitors, 1.23% kinase inhibitors and 1.2% nuclear receptor ligands. Band-cluster C (8820 molecules) contains 69.17% molecules of low or unspecified activities, 15.73% protease inhibitors, 6.09% family-A GPCR ligands, 2% reader inhibitors, 1.98% nuclear receptor ligands, 1.34% ion channel ligands, and 0.92% cytochrome P450 inhibitors. Within each band-cluster, molecules of the same bioactivity class are concentrated in several separated sub-regions. For instance, kinase inhibitors, GPCR ligands, and protease inhibitors in band-cluster A, B, and C are primarily in 9, 8 and 6 sub-regions respectively. Such banded distribution patterns were also found when molecules were inspected by Pfam or Interpro domain families.

**Unique combinations of sub-structural features and distinguished molecular scaffolds of individual band-cluster**

The concentrated distribution of selected bioactivity classes in individual band-clusters raises a question about what common sub-structural features are captured by the DAEs. Structural analysis of the molecules in several band-clusters showed that each band-cluster consists of molecules of unique combinations of sub-structural features. These involve combinations of three sub-structural elements (core hydrogen bond, core hydrophobic ring, linker between core elements) or minor structural variations of these elements. Figure 3 shows several representative molecules in band-cluster A, with their common sub-structural features highlighted by red (core hydrogen-bond element), blue (core hydrophobic ring element) and green (linker between core elements) colors. Supplementary Figure S2 and S3 show 298 and 295 molecules of band-cluster A and B respectively, with their common sub-structural features highlighted by the same color scheme.

In band-cluster A, the core hydrogen-bond element is of two framework groups. One is the Y-shaped N-N-N, N-O-N, N-S-N and O-N-O frameworks and their structural variations N-N-CN, N-N-CCN, N-N-C, N-O-CN, N-O-C, N-S-CN, and O-N-CO. Another is the line-shaped NC-N-CN framework and their structural variations NC-N-C, NC-N-CS, and NC-N-CCN. Multiple frameworks appear in a band-cluster because DAE-learned features are derived from weighted contributions of multiple sub-structures encoded in the Pubchem molecular fingerprints. These sub-structures and their variations may accord higher weight of importance and were subsequently captured by DAEs. The core hydrophobic ring element is of two frameworks, 6-member and 5-member carbon rings, plus different 1-heteroatom variations. Additional variations of these rings include 7-member carbon ring, multi-carbon single chain, and branched pair of short carbon chains or methyl-groups, which may contribute comparable hydrophobic interactions as a 6-member or 5-member carbon ring. The linker element is 1- to 4- carbon-bond chain.

In band-cluster B, the core hydrogen-bond element is of V-shaped NCN, NCO, OCO, SCS frameworks and their structural variations NCCN, NCC, NNCO, NCCO, and OCC. The core hydrophobic ring element is of the 6-member and 5-member ring frameworks, with each ring containing two-neighbor nitrogen heteroatoms (RNN), two non-neighbor nitrogen heteroatoms (RN2), 1-nitrogen heteroatom (RN), or all carbon atoms (R). The linker element is 1- to 4- carbon-bond chain connecting two core elements. Analysis of these and other band-clusters revealed that each band-cluster captures unique combinations of sub-structural elements fairly common within a band-cluster but largely distinguished among different band-clusters, consistent with the banding of different handwritten digits or different topics of Reuter newswire stories into different band-clusters (20, 21). While its members are characterized by unique combinations of the sub-structural features, each band-cluster primarily groups together molecules of diverse scaffolds or sub-structures with fewer overlaps with the scaffolds of other band-clusters. For instance, band-cluster A and B contain 37 and 99 molecular scaffolds respectively (Supplementary Table

S4), only six molecular scaffolds were found in both band-clusters (indoles, benzenoids, biphenyls, indolines, amino acids and pyrimidines).

**Relevance and distinction of DAE-captured sub-structural features with respect to the literature-reported pharmacophores and privileged structures.**

A question may be raised about the relevance of the DAE-captured sub-structural features to the selected bioactivities of the band-cluster. This question was probed by comparing the sub-structural features of band-cluster A (8.66% members are kinase inhibitors) with the literature-reported pharmacophores and privileged structures of kinase inhibitors(26, 28, 36), and the sub-structural features of band-cluster B (15.95% members are GPCR ligands) with the literature-reported privileged structures of GPCR ligands(37). Some sub-structural features of band-cluster A form the key frameworks of the literature-reported kinase-binding modes of kinase inhibitor drugs(28), pharmacophores of kinase frequent hitters(36) and privileged fragments of kinase inhibitors(26) (Figure 4 and Supplementary Figure S3). Specifically, these are the combinations of a core hydrogen-bond element (Y-shaped N-N-N, N-N-CN, N-S-N, N-O-C or L-shaped NC-N-CN, N-N-CN) and a core hydrophobic ring element. Some sub-structural features of band-cluster B also form the key frameworks of most of the literature-reported GPCR privileged structures(37) (Supplementary Figure S4). These are the combinations of a core hydrogen-bond element (V-shaped NCO, NCN, NCCO) and a core hydrophobic ring element (R and RN).

The connection of the DAE-captured sub-structural features to the frameworks of the selected bioactivities and the concentration of molecules of these bioactivities in a band-cluster is consistent with the reports that the kinase privileged fragments provide ~5-fold enrichment in kinase inhibitors (26). It is also consistent with the reports that some privileged molecular scaffolds enable activities against multiple target classes(25, 38). Nonetheless, the DAE-captured sub-structural elements differ from the literature-reported pharmacophores and privileged fragments in one aspect. DAEs capture the fundamental sub-structural elements and their structural variations whose combinations define the frameworks of the pharmacophores or privileged fragments. In contrast, the literature-reported kinase privileged structures are specific fragments such as the bisaryl-NH- linker fragments(26) and the bisphenylether scaffolds(38). Therefore, by learning the fundamental elements of structural frameworks, DAEs capture more variety of pharmacophores and privileged structures within individual band-cluster.

**Intra band-cluster distribution of bioactivity targets**

Given the banding of molecules of the selected bioactivity classes into selected band-clusters, a question arises about whether the molecules of specific target are further banded into selected sub-regions within a band-cluster. We analyzed the targets of the kinase inhibitors in band-cluster A (Supplementary Table S2) and the targets of the GPCR ligands in band-cluster B (Supplementary Table S3). While the molecules of the same target may be across multiple sub-regions, they tend to be banded in 1-3 selected sub-regions. In band-cluster A, the nine sub-regions populated by kinase inhibitors (Figure 2) are labeled as A1 to A9 from nearer the center of bioactive universe to the outward direction along the band-cluster. In A1, 9 of the 10 kinase inhibitors target MAPKAPK2. In A2, 15, 15, 14, 11, 11, 10, 9, 9 and 9 of the 80 kinase inhibitors (including multi-target inhibitors) target PDGFR□, VEGFR2, P38□, MAP2K1, EGFR PLK3, CLK4, FGFR1, and CDK2 respectively. In A3, 17 and 14 of the 89 kinase inhibitors target EGFR and CDK1. In A6, 17, 16 and 10 of the 30 kinase inhibitors target CDK1, GSK-3□□and CDK5. In A7, 15 and 9 of the 25 kinase inhibitors target AkT2 and PDPK1. In A8, 53, 34, 34, 22, 19,17, 16 and 16 of the 225 kinase inhibitors target VEGFR2, CDC7, PDGFR,□ALK, KIT, Chk1, MAPKAPK2, and RET respectively. In A9, 26, 16, and 16 of the 49 kinase inhibitors target PKC, PKA, and EGFR.

**Intra band-cluster distribution of sub-structural features and molecular scaffolds**

In addition to the banding of the molecules of specific target into selected sub-regions of a band-cluster, the DAE-captured sub-structural features and molecular scaffolds in a band-cluster are also banded into selected sub-regions. For instance, band-cluster A starts near the center of bioactive universe with sub-regions of higher concentrations of the Y-shaped N-S-N and N-S-CN frameworks. Going further out-ward along the band-cluster are sub-regions of higher concentrations of the line-shaped NC-N-CN, NC-N-CCN, NC-N-C, NC-N-CS frameworks and Y-shaped N-N-N, N-N-CN, N-N-CCN, N-N-C frameworks. Going toward the outer-edge are sub-regions of higher concentrations of the Y-shaped N-O-N, O-N-O, N-O-CN, N-O-C, and O-N-CO frameworks. Band-cluster B starts near the center of bioactive universe with sub-regions of higher concentrations of the RNN frameworks. Going outward along the band-cluster are sub-regions of higher concentrations of the RN2 and RN3 frameworks, followed by the sub-regions of higher concentrations of the R frames near the outer-edge.

The molecular scaffolds in sub-cluster A are also banded. A1 is enriched with N-phenylthioureas (8 molecules). A2 is populated by diazines, pyrimidines, diazanaphthalenes, thienopyrimidines, quinazolinamines, and benzodiazines (14, 14, 12, 12, 12 and 12 molecules). A3 is populated by benzimidazoles, diazanaphthalenes, benzodiazines, imidazopyridines and pyrimidines (11, 8, 8, 6, and 6 molecules). A6 is enriched with benzazepines and indoles (7 and 4 molecules). A7 is enriched with benzenoids, N-phenylureas and indolines (6, 5, and 3 molecules). A8 is populated by indoles, indolines, amino acids, pyrroles, carbazoles, pyrrolopyridines, and pyrrolocarbazoles (72, 34, 28, 24, 19, 15 and 13 molecules). A9 is populated by pyrrolocarbazoles, carbazoles, indolocarbazoles, and indoles (10 molecules each). A similar pattern of banded distribution of the molecular scaffolds were also found in band-cluster B.

**Concluding remarks**

DAEs showed superior capability in unsupervised clustering of big data into band-clusters of unique combinations of sub-component features but varying global features beyond data similarity. The deep learned combinations of sub-component features represent functionally or architecturally important intrinsic properties of the data in a band-cluster distinguished from other band-clusters. Analysis of the DAE-captured band-clusters of 1.39 million bioactive molecules showed that individual band-cluster groups together selected classes of bioactive molecules, whose structures are commonly composed of unique combinations of sub-structural features. These sub-structural features form the frameworks of the literature-reported pharmacophores and privileged fragments distinguished from other band-clusters. DAEs further band the molecules in a band-cluster into selected sub-regions enriched with higher concentration of the same bioactivity target and distinguished sets of molecular scaffolds.

These distinguished features of DAE-derived band-clusters are useful for grouping big data into functionally or architecturally relevant groups (clusters) and sub-groups (sun-regions). Our method may find extensive applications in big data clustering, stratification and knowledge discovery tasks in drug discovery(2), healthcare(3), biomedicine(4), biology(5), astronomy(39), economics(40) and other fields. Our method is based on the deep feature learning in a 3-dimensional latent space followed by visualization-facilitated analysis of data band-clusters. Although DAEs are capable of clustering big data into functionally and architecturally relevant groups and sub-groups, the learning in a 3-dimensioanl latent space instead of the usual higher dimensional latent space likely results in some degree of information loss, which may affect the quality and purity of the band-clusters and their sub-regions, as indicated by the scattered distribution of molecules of various bioactivity classes in individual band-clusters. Further efforts are needed for the development of bind-cluster extraction and evaluation methods in higher-dimensional latent space.

**Notes**

The authors declare no competing financial interests.

**ACKNOWLEDGMENT**

This work was funded by Shenzhen Municipal Government grants (JCYJ2016032416 3734374, JCYJ20170413113448742 and 20151030A0610001), Precision Medicine Project of the National Key Research and Development Plan of China (2016YFC 0902200) and the Innovation Project on Industrial Generic Key Technologies of Chongqing (cstc2015zdcy-ztzx120003).

14. Eulenberg P, *et al.* (2017) Reconstructing cell cycle and disease progression using deep learning. *Nat Commun* 8(1):463.

15. Han Z, *et al.* (2017) Mesh Convolutional Restricted Boltzmann Machines for Unsupervised Learning of Features With Structure Preservation on 3-D Meshes. *IEEE Trans Neural Netw Learn Syst* 28(10):2268-2281.

16. Young JD, Cai C, & Lu X (2017) Unsupervised deep learning reveals prognostically relevant subtypes of glioblastoma. *BMC Bioinformatics* 18(Suppl 11):381.

17. Kallenberg M, *et al.* (2016) Unsupervised Deep Learning Applied to Breast Density Segmentation and Mammographic Risk Scoring. *IEEE Trans Med Imaging* 35(5):1322-1331.

18. Silver D, *et al.* (2017) Mastering the game of Go without human knowledge. *Nature* 550(7676):354-359.

19. Gebru T, *et al.* (2017) Using deep learning and Google Street View to estimate the demographic makeup of neighborhoods across the United States. *Proc Natl Acad Sci U S A* 114(50):13108-13113.

20. Hinton GE & Salakhutdinov RR (2006) Reducing the dimensionality of data with neural networks. *Science* 313(5786):504-507.

21. Shin HC, Orton MR, Collins DJ, Doran SJ, & Leach MO (2013) Stacked autoencoders for unsupervised feature learning and multiple organ detection in a pilot study using 4D patient data. *IEEE Trans Pattern Anal Mach Intell* 35(8):1930-1943.

22. Hosseini-Asl E, Zurada JM, & Nasraoui O (2016) Deep Learning of Part-Based Representation of Data Using Sparse Autoencoders With Nonnegativity Constraints. *IEEE Trans Neural Netw Learn Syst* 27(12):2486-2498.

23. Gomez-Bombarelli R, *et al.* (2018) Automatic Chemical Design Using a Data-Driven Continuous Representation of Molecules. *ACS Cent Sci* 4(2):268-276.
15

**FIGURE LEGENDS**

**Figure 1.** The band-cluster landscapes of 1.39 million Chembl bioactive molecules in the DAE-captured 3-dimentional latent chemical space, displayed on DeepChemscape. Top view: molecules are represented by individual spheres in default grey color, except 10 band-clusters highlighted by other colors. Two, five and three of these 10 band-clusters are populated by kinase inhibitors (blue, lime), GPCR ligands (beige, fuchsia, pink, teal, purple), and protease inhibitors (green, yellow). The red band-cluster is dominated by molecules of unspecified bioactivity. Bottom view: 10 colored band-clusters alone.

**Figure 2.** Three band-clusters populated by molecules of selected bioactive classes. The top, middle and bottom band-cluster A, B, and C populated by kinase inhibitors (blue), GPCR ligands (orange), and protease inhibitors (green) respectively. The other bioactivity classes with significant populations are nuclear receptor ligands (purple) and voltage-gated ion channel inhibitors (pink).The center of bioactive universe is at the left bottom corner of each sub-figure.

**Figure 3.** The structures of the selected molecules of band-cluster A in Figure 2. Their common substructural features highlighted by red (core hydrogen-bond element), blue (core hydrophobic ring element) and green (linker between core elements) colors.

**Figure 4.** The key pharmacophoric elements of the binding modes of kinase inhibitor drugs reported in literature. The core pharmacophoric elements highlighted by red (core hydrogen-bond element), blue (core hydrophobic ring element) and green (linker between core elements) colors.